\documentstyle[12pt,epsf,graphicx]{article}
\def\href#1#2{#2}   % Ignore hypertext links in UTPHYS.BST
\newif\ifdraft
\draftfalse	  % \drafttrue for draft, \draftfalse for final version
\reversemarginpar   % Use the left margin
\makeatletter
\let\mlabel=\label
\let\adkendequation=\endequation%
\def\endequation{\adkendequation\adklabel\global\@ignoretrue}
\let\adkendeqnarray=\endeqnarray%
\def\endeqnarray{\adkendeqnarray\adklabel\global\@ignoretrue}
\newbox\marglabbox
\def\adklabel{\ifvoid\marglabbox\else\marginpar{\unhbox\marglabbox}\fi}
\def\label#1{\ifdraft\ifmmode%
  \global\setbox\marglabbox=\hbox{\hfill\fbox{\tiny\verb*~#1~}}%
  \else\ifinner\else\marginpar{\hfill\fbox{\tiny\verb*~#1~}}%
  \fi\fi\fi \mlabel{#1}}
\makeatother
\ifdraft%
\fi
\font\twelvebb=msbm12
\font\tenbb=msbm10
\font\sevenbb=msbm7
  \newfam\bbfam
  
  \textfont\bbfam=\twelvebb
  \scriptfont\bbfam=\tenbb
  \scriptscriptfont\bbfam=\sevenbb
\font\twelveeusm=eusm10 scaled 1200
\font\teneusm=eusm10
  \newfam\eusmfam
  
  \textfont\eusmfam=\twelveeusm
  \scriptfont\eusmfam=\teneusm
  \scriptscriptfont\eusmfam=\scriptfont\eusmfam
\font\twelvefrak=eufm10 scaled 1200
\font\tenfrak=eufm10
  \newfam\frakfam
  
  \textfont\frakfam=\twelvefrak
  \scriptfont\frakfam=\tenfrak
  \scriptscriptfont\frakfam=\scriptfont\frakfam
\def\sqr#1#2{{\vcenter{\hrule height.#2pt
   \hbox{\vrule width.#2pt height#1pt \kern#1pt
      \vrule width.#2pt}
   \hrule height.#2pt}}}

\def\bsqr#1#2{{\vrule width #1pt height#2pt}}
\def\bsquare{{\mathchoice\bsqr66\bsqr66\bsqr33\bsqr33}}
\def\badbreak{\penalty1000}

\newcommand{\cE}{{\cal E}}                  % cal-E
\newcommand{\cR}{{\cal R}}                  % cal-K
\begin{document}

\begin{center}
{\Large{\bf The Reality of the Fundamental Topological}}\\
\vspace{.1in}
{\Large{\bf Structure in the QCD Vacuum}}\\
\vspace*{.35in}
{\large{Andrei Alexandru$^1$, Ivan Horv\'ath$^1$ and Jianbo Zhang$^2$}}\\
\vspace*{.15in}
$^1$Department of Physics and Astronomy, University of Kentucky, Lexington, KY 40506\\
$^2$CSSM and Department of Physics, University of Adelaide, Adelaide, SA 5005, Australia

\vspace*{0.2in}
{\large{Jun 13 2005}}

\end{center}

\vspace*{0.10in}

\begin{abstract}
  \noindent
  Long-range order of a specific kind has recently been found directly in configurations 
  dominating the regularized QCD path integral. In particular, a low-dimensional
  global structure was identified in typical space-time distributions of topological 
  charge defined via the overlap Dirac matrix. The presence of the order has been 
  concluded from the fact that the structure disappears after random permutation 
  of position coordinates in measured densities. Here we complete the argument for 
  the reality of this structure (namely the conjecture that its existence is a consequence 
  of QCD dynamics and not an artifact of the overlap-based definition of lattice 
  topological field) by showing that the structure ceases to exist after randomizing 
  the space-time coordinates of the underlying gauge field. This implies that the long-range 
  order present in the overlap-based topological density is indeed a manifestation of QCD 
  vacuum, and that the notion of the {\em fundamental structure} (structure involving 
  relevant features at all scales) is viable.

\end{abstract}

Numerical lattice gauge theory represents a powerful tool for obtaining quantitative
predictions of QCD from first principles. However, apart from extracting physical 
quantities via numerical simulation, lattice theory is also being used in attempts 
to decipher the nature of the QCD vacuum. This is quite natural since the latter problem 
is frequently approached via the {\em hypothesis} that there exist certain well-defined 
objects (``structure'') that dominate the behavior of typical gauge configurations 
contributing to Euclidean QCD path integral. Specific properties of such objects are then 
expected to encode the mechanism QCD uses to induce confinement, spontaneous chiral symmetry 
breaking and other effects. If one accepts this logic, then a straightforward approach 
to the problem of QCD vacuum is to search for a well-defined structure in configurations 
dominating the evaluation of physical observables in regularized theory, i.e. in equilibrium 
Monte Carlo configurations of finite lattice systems. Unfortunately, for a long time, such 
a direct approach has not been fruitful since no obvious structure has been observed neither 
in unmodified equilibrium configurations of the fundamental gauge field nor in the 
configurations of relevant composite fields derived from it. In fact, the configurations 
in accessible ensembles appeared to be more or less structureless. 

The usual explanation of this fact is rather vague and involves variations 
on the proposition that large entropy of fluctuations at the scale of the cutoff 
obscures any ordered structure that might be present (the {\em ``entropy problem''}). 
However, when making this argument, one should (at least mentally) distinguish two possible 
origins of ultraviolet fluctuations in question.
(i) Given a lattice cutoff $\Lambda =1/a$ there are legitimate {\em physical} QCD 
fluctuations at this scale. There is no reason to expect that this part of fluctuations 
at the scale of the lattice cutoff should be structureless. Indeed, it would be quite
unnatural if the hypothesis of the structure applied only at long distances in the theory with 
non-trivial ultraviolet behavior.
(ii) There are unphysical fluctuations that appear only as artifacts of a field-theoretic
description of strong interactions. Such fluctuations are regularization--dependent and can 
be entirely structureless. From this point of view, the entropy problem is not really the 
problem of large entropy associated with ultraviolet fluctuations. It is rather the problem of 
large contribution of artifacts relative to physically relevant fluctuations at the scale of 
the cutoff. The severity of the issue can thus strongly depend on the lattice action used 
to define QCD, as well as on the choice of lattice operators for composite fields 
of interest.

Adopting the above (heuristic) logic as a starting point, the issue of identifying the QCD 
vacuum structure in unmodified equilibrium configurations has been revisited in 
Refs.~\cite{Hor03A,Hor05A}. In particular, the configurations of lattice topological field 
defined via Dirac kernel of exactly chiral lattice fermions~\cite{Has98A,NarNeu95} have
been computed on unmodified equilibrium gauge backgrounds of pure-glue QCD. The underlying
expectation was that the structure, not apparent in the gauge field itself, could become 
visible in this particular composite field. This was motivated by the fact that the corresponding 
topological charge density (TChD) operator is constructed in a very different manner than 
the standard lattice operators, which translates into beautiful continuum-like behavior of 
this composite field already at the regularized level~\cite{Giu02A,GiuRosTes04,Lusch04}. 
At the same time, and more importantly for our purposes, it is expected that the lattice 
operators in this class are necessarily non-ultralocal (but still local), similarly 
to non-ultralocality of the associated fermionic action~\cite{nonultr}. This should soften 
the impact of ultraviolet gauge fluctuations on the topological field. Moreover, such 
{\em chiral smoothing}~\cite{Hor02A} is expected to be very efficient in eliminating 
the structureless artifacts-related ultraviolet fluctuations, while still preserving 
the physical short-distance fluctuations~\cite{Hor03Apr}, thus providing the window of 
opportunity for avoiding the entropy problem.

The numerical experiments of Refs.~\cite{Hor03A,Hor05A} indeed revealed the existence of 
a non-trivial space-time structure in the overlap-based topological field, contrary to 
the absence of an observable order when standard naive operators are used.~\footnote{Similar 
results have subsequently been reported also in the case of 2-d CP(N-1) models~\cite{Tha04}.}
One particular manifestation of the structure is that the topological charge in typical 
configurations organizes into two sign-coherent locally low-dimensional 
{\em ``sheets''}.~\footnote{The notion of strictly low-dimensional structure has recently been 
invoked also in the work using projected gauge fields associated with gauge-fixing 
procedures~\cite{Zakh}. It remains to be seen whether there exists a connection to 
the topological structure on equilibrium backgrounds.} The sheets can be ``tiled'' with 
3-d sign-coherent elementary cubes connected via 2-d faces but not with 4-d coherent cubes. 
In fact, the fraction of space-time occupied by connected sign-coherent regions built of 
4-d hypercubes scales to zero in the continuum limit, thus excluding the possibility of 
the coherence on smooth 4-d manifolds. The double-sheet structure is global in the sense 
that each sheet spreads over largest possible distances. It contains a global connected  
substructure -- the {\em ``skeleton''} -- consisting of approximately 1-d filaments of 
strong fields. Both the sheets and the skeleton fill a macroscopic (non-zero and, 
in fact, large) fraction of space-time and their geometric nature is analogous to that 
of the Peano's curve, i.e. a structure with local attributes of a low-dimensional object 
but still filling the underlying higher-dimensional space. An important aspect 
of the order in topological field is that the two sheets, as well as the oppositely charged 
parts of the skeleton, are embedded in space-time in a mutually correlated manner so as 
to yield a negative two-point function of TChD~\cite{Hor03A,SeSt,Hor05B}.

The details of the vacuum structure described above, while expected to be relevant for future 
understanding of the role of vacuum in strong-interaction physics, mainly serve here 
as an unbiased evidence for the basic conceptual point put forward in Ref.~\cite{Hor03A}. 
In particular, they support the proposition that there exists a {\em fundamental structure} 
(structure involving features at all scales) in gauge configurations dominating the QCD path 
integral. This structure can be identified and studied directly in lattice-regularized ensembles, 
and its existence is a direct consequence of QCD dynamics. The above conclusion, if established 
beyond doubt, is quite far-reaching since, in both conceptual and practical sense, it forms 
a basis for possible understanding of the QCD vacuum in the Euclidean path-integral formalism.
To demonstrate the existence of the fundamental structure one has to show that there exists 
a measurable excess of order in any configuration typical of QCD ensemble relative to 
a configuration constructed randomly. In Ref.~\cite{Hor03A} this has been addressed at 
the level of the topological field itself. More specifically, it was shown that after 
randomly permuting the space-time coordinates of topological densities measured in typical 
gauge backgrounds, the ordered structure described in the previous paragraph ceases to exist. 
In particular, the sheets built from 3-d coherent lattice hypercubes disappear after such 
random reshuffling of TChD in a given configuration. While indirect, this represents 
a strong argument supporting the reality of the fundamental structure in QCD vacuum.

The reason why the above argument is indirect is that it compares the degree of order in the
composite field evaluated on equilibrium gauge background relative to the situation in 
a disordered composite field. A direct approach should compare the former relative to 
the situation in composite field evaluated on a disordered gauge background. 
To see the difference between the two procedures at the technical level more precisely, 
let us denote collectively $U^{QCD} \equiv \{\, U(x,\mu) \,\}$ an equilibrium QCD gauge 
configuration and $q^{QCD} \equiv \{\, q(x) \,\}$ the associated configuration of TChD. 
If $x \longrightarrow p_S(x)$ represents a random permutation of the (scalar) space-time
coordinates then the operation of randomizing the configuration of TChD corresponds to
\begin{equation}
   q^{QCD} \equiv \{\, q(x) = q(x,U^{QCD}) \,\}   \quad\longrightarrow\quad
   q^{R}  \equiv \{\, q^{R}(x) = q(p_S(x),U^{QCD}) \,\}
   \label{eq:5}
\end{equation}
At the same time, if $(x,\mu) \longrightarrow p_V(x,\mu)$ represents a random 
permutation of link (vector) space-time coordinates, then the randomization of gauge 
configuration proceeds via
\begin{equation}
   U^{QCD} \equiv \{\, U(x,\mu) \,\}   \quad\longrightarrow\quad
   U^{R}  \equiv \{\, U^{R}(x,\mu) = U(p_V(x,\mu)) \,\}
   \label{eq:10}
\end{equation}
and the associated configuration of TChD is then
\begin{equation}
   q^{QCD} \equiv \{\, q(x) = q(x,U^{QCD}) \,\}   \quad\longrightarrow\quad
   q^{RU}  \equiv \{\, q^{RU}(x) = q(x,U^{R}) \,\}
   \label{eq:15}
\end{equation}
If there is an excess of structure in $q^{QCD}$ relative to $q^{RU}$, then it can be directly 
ascribed to the underlying order in the gauge field induced by QCD dynamics. On the other hand, 
if this is not the case then the structure observed in TChD is mainly induced by non-ultralocal
nature of the overlap-based TChD operator. In other words, the structure in $q^{QCD}$ would arise 
due to artificial amplification of random seeds of coherence, and would not be a manifestation 
of QCD dynamics. 

In this work we perform a numerical experiment to address this issue. In particular, we study
the effect of $q^{QCD} \longrightarrow q^{RU}$ on the double-sheet structure in the same
manner as was done for $q^{QCD} \longrightarrow q^{R}$ in Ref.~\cite{Hor03A}. We computed 
both $q^{QCD}$ and $q^{RU}$ for five independent $8^4$ equilibrium configurations of Iwasaki 
gauge action at lattice spacing $a=0.165$~fm determined from string tension.~\footnote{These 
are actually the first five configurations of ensemble $\cE_1$ of Ref.~\cite{Hor05B}.} 
Topological densities were computed using~\cite{Has98A}
\begin{equation}
   q(x) \;=\; \frac{1}{2\rho} \, \mbox{\rm tr} \,\gamma_5 \, D_{x,x} \;\equiv\;  
             -\mbox{\rm tr} \,\gamma_5 \, (1 - \frac{1}{2\rho}D_{x,x})
   \label{eq:20}  
\end{equation}  
where $D$ is the overlap Dirac operator~\cite{Neu98BA} based on the Wilson-Dirac 
kernel with mass $-\rho$. Numerical results presented here were obtained at the value
$\rho=1.368$ ($\kappa=0.19$). Details of the numerical implementation for overlap 
matrix--vector operation needed to evaluate $q(x)$ can be found in Ref.~\cite{Chen03}. 
We wish to point out that while the number of configurations used in this study might seem 
small, it was found in Refs.~\cite{Hor03A,Hor05A} that the qualitative (and even quantitative) 
behavior of the observed structure is very robust and changes very little from one 
configuration to another. This is in fact expected if typical configurations are indeed 
dominated by a specific kind of space-time structure. Consequently, a qualitative conclusion 
(such as one sought here) can be made from just a handful of configurations. 

   \begin{figure}
   \begin{center}
     \vskip -0.40in
     \centerline{
     \hskip -0.00in
     \includegraphics[height=21.5truecm,angle=0]
                     {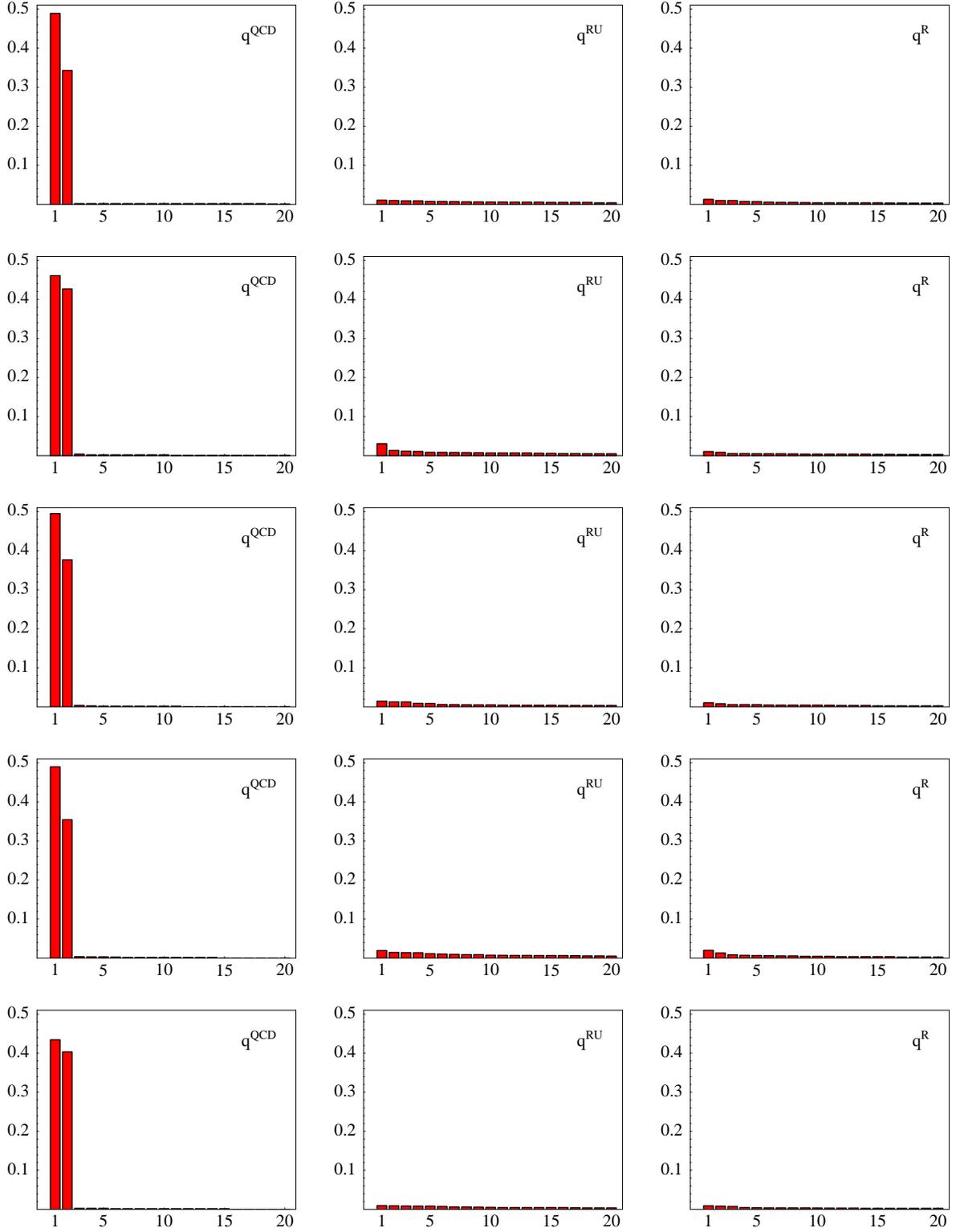}
     }
     \vskip -0.15in
     \caption{Fractions $f_k$ associated with maximal connected structures $\cR^3_k$
              ordered by size (decreasing $f_k$) are plotted against $k$. Each row corresponds 
              to an individual configuration with columns representing $q^{QCD}$, $q^{RU}$ and 
              $q^R$ respectively.}
     \vskip -0.4in 
     \label{str_indv:fig}
   \end{center}
   \end{figure}

   \begin{figure}[t]
   \begin{center}
     \vskip -0.10in
     \centerline{
     \includegraphics[width=7.8truecm,angle=0]{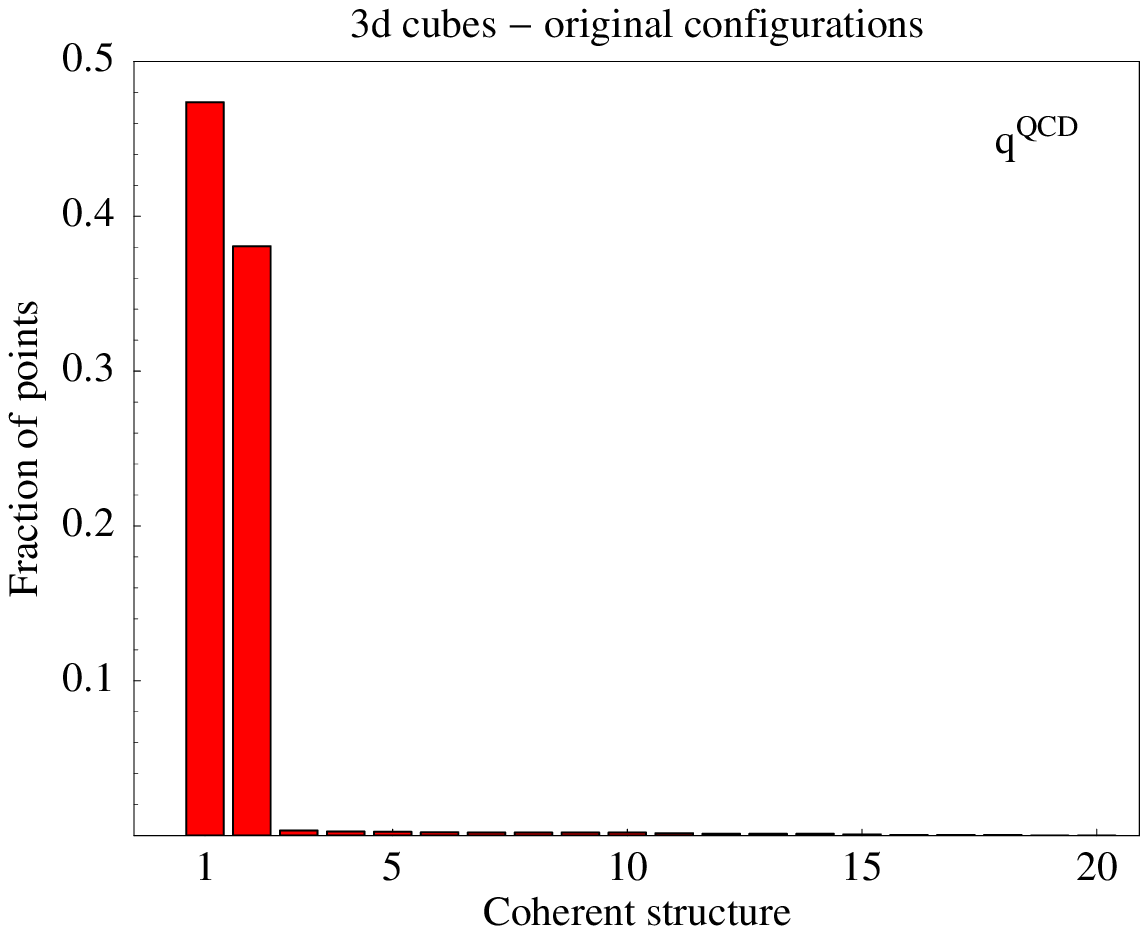}
     \includegraphics[width=7.8truecm,angle=0]{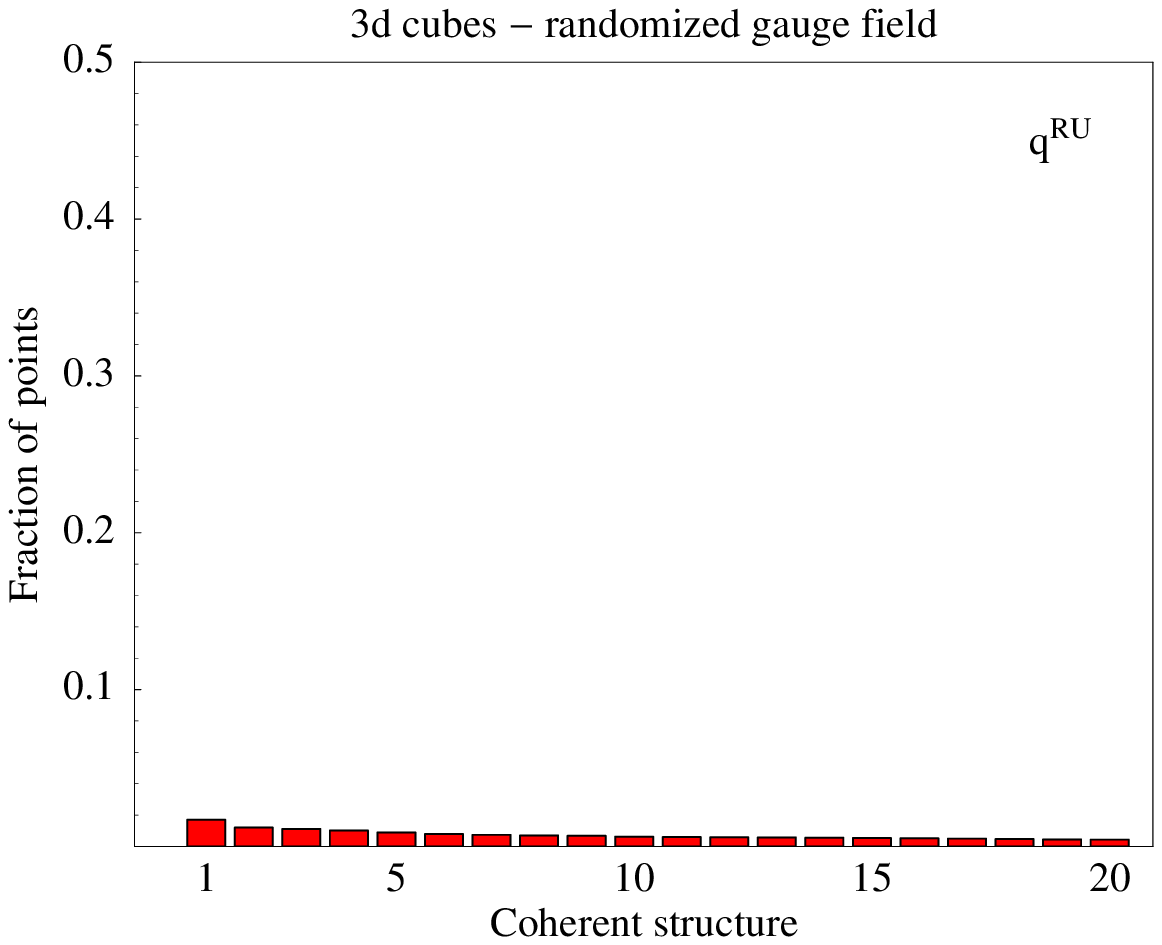}
     }
     \vskip 0.20in
     \centerline{
     \includegraphics[width=7.8truecm,angle=0]{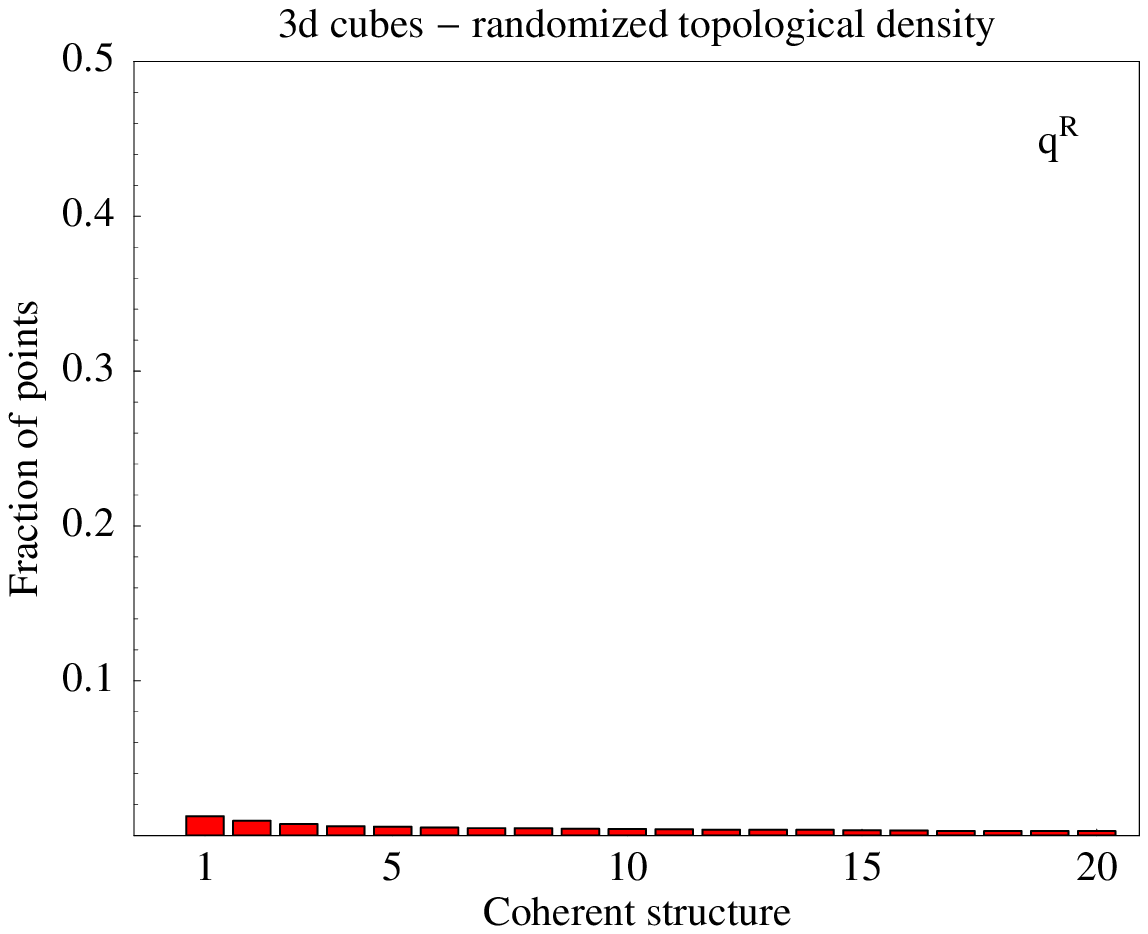}
     \includegraphics[width=7.8truecm,angle=0]{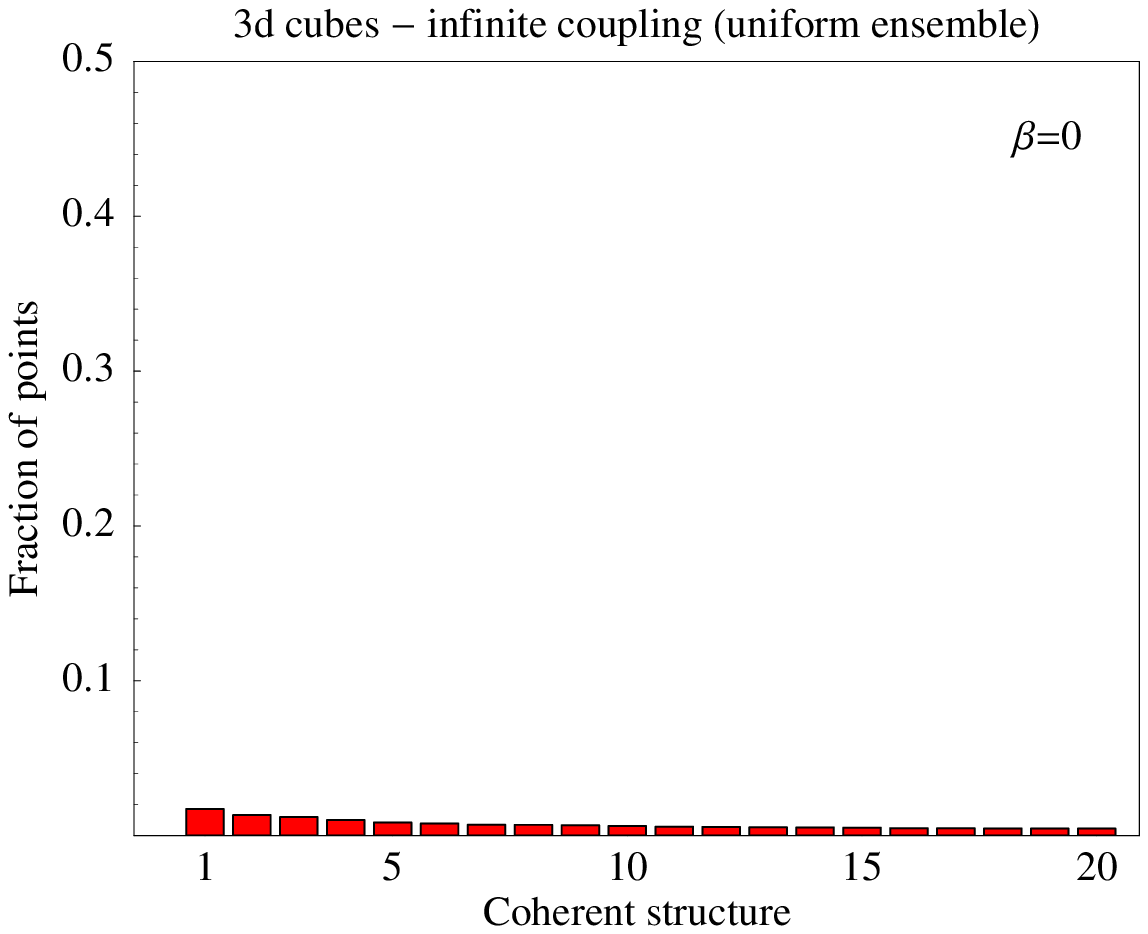}
     }
     \caption{The average fractions $f_k$ are plotted against $k$ for $q^{QCD}$ (top left),
              $q^{RU}$ (top right), $q^{R}$ (bottom left) and for uniform ($\beta=0$) 
              ensemble (bottom right).}
     \label{str_av:fig}
     \vskip -0.35in
   \end{center}
   \end{figure}

For a given configuration of TChD we follow the procedure of Ref.~\cite{Hor03A} and determine 
all maximal connected sign-coherent regions $\cR^3_k, \,k=1,\ldots ,K_3$ that can be built 
from elementary 3-d cubes. We emphasize that the cubes are connected via 2-d faces so that a 
consistent lattice 3-d hypersurface (individual ``structure'') is defined by any $\cR^3_k$.
Let $N(\Gamma)$ denote the number of points in arbitrary subset $\Gamma$ of discretized
space-time $\Omega$. Then the fraction $f_k$ of space-time occupied by $\cR^3_k$ is given 
by $f_k\equiv N(\cR^3_k)/N(\Omega)$. In what follows we order the structures by decreasing
fraction, i.e. we choose the enumeration such that $f_k \ge f_{k+1}$ for all $k\le K_3-1$.
The resulting sequences $f_k$ (for $k \le 20$) are plotted in Fig.~\ref{str_indv:fig} with 
each row representing a situation in a given individual configuration. The first column shows 
$f_k$ for $q^{QCD}$ with the double-sheet structure appearing in each configuration via 
dominance of $f_1$ and $f_2$.~\footnote{The results of Ref.~\cite{Hor03A} indicate that both 
$f_1$ and $f_2$ have a finite continuum limit with the combined double-sheet occupying about 
70--80\% of space-time. On the other hand, the fractions $f_k$ for $k\ge 3$ (the ``fragments'') 
scale to zero in the continuum limit.} In the second column the corresponding fractions are 
shown for $q^{RU}$, i.e. for randomized gauge field defined in Eq.~(\ref{eq:15}). The key point
of this work is that, as clearly seen from the plots, the double-sheet structure disappears
when $q^{QCD} \longrightarrow q^{RU}$. Indeed, only fragmented pieces of coherence survive
the random reshuffling of space-time coordinates of the gauge field. In fact, the situation 
is very similar to randomization of topological density ($q^{QCD} \longrightarrow q^{R}$) 
as shown in the right column. 

In Fig.~\ref{str_av:fig} we plot the configuration averages from the data shown 
in Fig.~\ref{str_indv:fig}. In addition, we have included here (lower right plot) the result 
from the same structure analysis for five configurations of $8^4$ lattice generated at infinitely 
strong coupling ($\beta=0$). This represents another form of comparison between the structure 
in QCD and the situation for disordered gauge field. Indeed, an equilibrium configuration from 
this ensemble can be generated by independently choosing each link from a uniform distribution. 
As can be clearly seen from the plots, the result is again completely analogous to that 
of $q^{RU}$ and $q^R$. We thus conclude that the double-sheet structure exists due to the 
underlying order in gauge configurations dominating the QCD path integral, and is indeed 
a manifestation of QCD vacuum.

To summarize, the goal of this work was to provide additional evidence for the proposition
that there exists a {\em fundamental structure} (space-time order involving all scales) in 
configurations dominating the QCD path integral~\cite{Hor03A}. In particular, it was shown 
that a particular manifestation of it, namely the global double-sheet structure observed in 
the configurations of overlap-based TChD, only exists when appropriate local correlations 
(local QCD interaction) governs the ensemble. This long-range structure disappears when 
the space-time correlation among gauge variables is turned off. From this we conclude that 
the double-sheet structure (as well as the underlying gauge structure inducing it) is {\em real} 
in the sense that it exists as a consequence of QCD dynamics. We wish to emphasize two points 
that we associate with this finding (see also~\cite{Hor03A}).
{\em (i)} The observation of structure directly in equilibrium configurations via
a physically relevant composite field puts the {\em hypothesis} of the structure, and 
with it the approach to QCD vacuum via Euclidean path integral formalism, on a firmer 
ground. It also suggests that a direct systematic approach using ensembles of lattice 
QCD is a viable (and perhaps optimal) avenue to study the problem of QCD vacuum. 
{\em (ii)} Our approach leads us to the conclusion that vacuum structure should not be
viewed as a purely low-energy concept. Indeed, the structure observed in 
Refs.~\cite{Hor03A,Hor05A} involves space-time features at all scales (upon taking 
the continuum limit), and should have manifestations relevant for physics at arbitrary 
energy. The behavior of the structure at a given fixed scale can be studied in case 
of TChD via {\em effective densities}~\cite{Hor02B} obtained by the means of Dirac eigenmode 
expansion. We emphasize that the qualitative novelty here consists in the view that 
``understanding'' the vacuum structure crucially involves an insight into how the structure 
changes across all scales. 

\bigskip\medskip
\noindent
{\bf Acknowledgments:} 
We are indebted to Terry Draper and Keh-Fei Liu for a useful feedback on the manuscript. 
This work was supported in part by U.S. Department of Energy under grants 
DE-FG05-84ER40154 and DE-FG02-95ER40907.

\end{document}
\bye